\documentclass[epj]{svjour}
\usepackage{graphicx}

\newcommand{\beq}{\begin{equation}} 
\newcommand{\eeq}{\end{equation}}
\newcommand{\bea}{\begin{eqnarray}} 
\newcommand{\eea}{\end{eqnarray}}

\begin{document}

\title{Velocity filtration and temperature inversion in a system with long-range interactions} 
\author{Lapo Casetti\inst{1,2}\thanks{\email{lapo.casetti@unifi.it}} \and Shamik Gupta\inst{3}\thanks{\email{shamikg1@gmail.com}}}
\titlerunning{Velocity filtration and temperature inversion in a system with long-range interactions}
\authorrunning{L.\ Casetti and S.\ Gupta}
\institute{Dipartimento di Fisica e Astronomia and CSDC,
Universit\`a di Firenze, and INFN, Sezione di Firenze, via G.\ Sansone 1, I-50019 Sesto Fiorentino (FI), Italy \and
INAF-Osservatorio Astrofisico di Arcetri, Largo E.\ Fermi 5, I-50125 Firenze, Italy \and  
Laboratoire de Physique Th\'{e}orique et Mod\`{e}les
Statistiques (CNRS UMR 8626), Universit\'{e} Paris-Sud, Orsay, France
}
\date{Received: date / Revised version: date}

\abstract{Temperature inversion due to velocity filtration, a mechanism originally
proposed to explain the heating of the solar corona, is demonstrated to
occur also in a simple paradigmatic model with long-range interactions, the Hamiltonian mean-field model. Using molecular dynamics simulations, we show that when the system settles
into an inhomogeneous quasi-stationary state in which the velocity distribution
has suprathermal tails, the temperature and density profiles are
anticorrelated: denser parts of the system are colder than dilute ones. We argue that this may be a generic property of long-range interacting systems.
}

\maketitle

\section{Introduction}
Velocity filtration is a dynamical phenomenon originally suggested by
Scudder \cite{Scudder:apj1992a,Scudder:apj1992b,Scudder:apj1994} to
explain temperature inversion in inhomogeneous
plasmas like the outer atmosphere of the Sun, i.e., the solar corona\footnote{This phenomenon occurs also in the Io plasma torus around Jupiter and might have the same origin, as suggested in Ref.\ \protect\cite{MeyerVernetMoncuquetHoang:icarus1995}.}. By temperature inversion
we mean that the temperature grows while the density decreases with
increasing distance from the
photosphere, so that temperature and density
profiles are anticorrelated: Temperatures of order $10^6$ K are
attained in the corona \cite{GolubPasachoff:book}. Despite recent
advances \cite{Aschwanden:book}, the question of what heats the solar
corona remains one of the most important open problems in astrophysics
\cite{Klimchuk:sp2006}. Although velocity filtration itself might not be
the complete solution of the problem
\cite{Klimchuk:sp2006,Anderson:apj1994,LandiPantellini:aa2001}, it
provides a simple explanation of how such a counterintuitive phenomenon may occur in general.

This paper does not address velocity filtration and temperature
inversion in its original astrophysical context. It rather aims at
showing that this phenomenon is very general and may occur in situations
quite far from that originally envisaged by Scudder. In particular,
temperature inversion due to velocity filtration may occur in a system
with long-range interactions\footnote{Interactions are called
long-range when they decay asymptotically with the interparticle distance $r$ as
$r^{-\alpha}$ with $0 \leq \alpha < d$ in $d$ dimensions. See e.g.\
Refs.\ \cite{CampaEtAl:physrep,BouchetGuptaMukamel:physicaA2010}.} in a
quasi-stationary state (QSS), provided the particle distribution in the
QSS is inhomogeneous. 
QSSs are out-of-equilibrium states that occur during relaxation of an isolated long-range interacting
system to equilibrium. Due to a fast ``violent'' relaxation on times of
$O(1)$, a generic initial condition reaches a QSS. These states persist
for very long times, diverging with the number of degrees of freedom $N$ as $N^\alpha$ with $\alpha \geq 1$. Then, finite-$N$ effects drive the system to equilibrium.
The lifetime of a QSS is thus effectively infinite in the thermodynamic limit $N\to\infty$ \cite{CampaEtAl:physrep,BouchetGuptaMukamel:physicaA2010}. 

The system we shall study in this paper is a paradigmatic model with
long-range interactions, the so-called Hamiltonian Mean Field (HMF)
model \cite{AntoniRuffo:pre1995}. The model can be seen either as a
system of globally coupled point particles moving on a circle or as
a system of fully coupled (mean-field) $XY$ spins, with Hamiltonian   
\beq
\mathcal{H} = \frac{1}{2}\sum_{i=1}^N p_i^2 + \frac{J}{N}{\sum_{i=1}^N \sum_{j< i}^N} \left[ 1 - \cos \left(\vartheta_i - \vartheta_j \right) \right] - h\sum_{i=1}^N \cos \vartheta_i \, .
\label{hmf}
\eeq
Here $\vartheta_i\in (0,2\pi]$ is the angular coordinate of the $i$th
particle ($i = 1,\ldots,N$) on the
circle, while $p_i$ is the conjugated momentum. 
In the following, we shall consider two cases: (i) the 
antiferromagnetic (AF) case ($J < 0$), with an external field $h > 0$, and
(ii) the ferromagnetic (F) case ($J > 0$), without the external field, i.e., $h =
0$. In the AF case, the system settles into an inhomogeneous QSS due to the
action of the external field that forces the particles to cluster around
$\vartheta = 0$. In the F case, a clustered QSS is attained via a
spontaneous breaking of the $O(2)$ symmetry by the
attractive interaction, provided the total energy is small enough.
The HMF is a toy model not capable of accurately
describing any real physical system. However, the cosine interaction in
Eq.\ (\ref{hmf}) is the first mode of a Fourier expansion of a Coulomb or gravitational potential
energy. Hence, the AF case may be considered as a simplified model of a
one-component plasma in an external field that promotes clustering, while the F case
may be seen as a simplified model of a self-gravitating system
\cite{AntoniRuffo:pre1995,ChavanisVattevilleBouchet:epjb2005}. Moreover, a system of self-gravitating particles with a short-distance regularisation and constrained on a ring
\cite{SotaEtAl:pre2001} is equivalent to the HMF model in the limit of large regularisation scale \cite{TatekawaEtAl:pre2005,jstat2010}. 

By means of molecular dynamics (MD) simulations, we shall demonstrate
that temperature inversion occurs in both the AF and the F case, for
suitable classes of initial conditions with suprathermal velocity
distributions (i.e., tails ``fatter'' than those of a Maxwellian),
provided the QSS the system settles into
is clustered. Before describing the numerical simulations and discussing
our results, let us briefly recall the original argument due to Scudder
\cite{Scudder:apj1992a}, and discuss why it should also hold in our situation.

\section{Velocity filtration and temperature inversion}
\begin{figure}
\centerline{\includegraphics[width=85mm]{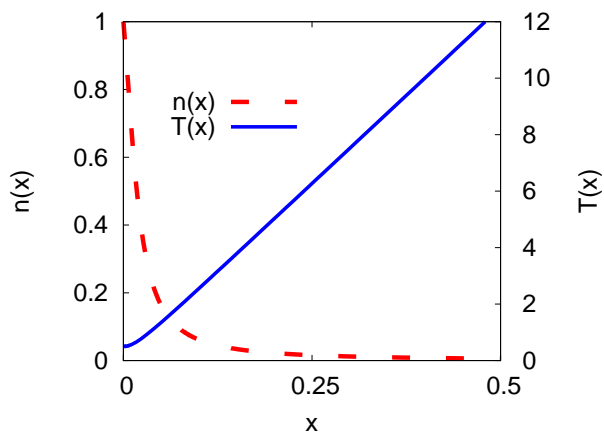}}
\caption{(Color online) Temperature (blue solid line) and density (red dashed line) profiles calculated for noninteracting particles in a potential $\psi(x) = 25\,x$ with $f_0(p) \propto p^{-4}$ as given by Eq.\ (\ref{f04}).}
\label{fig_scudder_calculation}
\end{figure}

Let us consider a system of noninteracting particles moving in one
dimension on a semi-infinite line $x \geq 0$ in presence of an
external potential $\psi(x)$, such that $\psi'(x) > 0$.
A typical case is that of an atmosphere, where $x$ is the height above the surface and the potential $\psi(x)$ is the gravitational one. The one-particle distribution function $f(x,p,t)$ obeys the Vlasov equation
\beq
\frac{\partial f}{\partial t} + p \frac{\partial f}{\partial x} -
\frac{d\psi}{dx}\frac{\partial f}{\partial p} = 0\, ,
\label{vlasov}
\eeq 
with a given stationary boundary condition at $x = 0$, i.e.,
\beq
f_0(p) \equiv f(0,p,t)\, .
\eeq
The Vlasov equation (\ref{vlasov}) expresses energy conservation for
each particle, so that due to the potential $\psi(x)$, there
will be a ``velocity filtration'' effect. Namely, only those particles whose kinetic energy $k$ at $x = 0$ is sufficiently large to overcome the potential barrier $\Delta\psi(x') = \psi(x') - \psi(0)$  will reach the position $x'$ where their kinetic energy will be $k' = k - \Delta\psi(x')$. This is the dynamical origin of the density
\beq
n(x) = \int_{-\infty}^{\infty} dp\, f(x,p)
\eeq
being a decreasing function of $x$. Consider a Maxwellian boundary condition,
\beq
f_0^{\rm M}(p) = \frac{n_0}{\left(2\pi T_0 \right)^{1/2}} \exp\left(-\frac{p^2}{2T_0} \right)\, ,
\eeq
where $n_0 = n(0)$, and for ease of notation, we take particles of unit
mass and set the Boltzmann constant $k_B$ to unity. Then, the
stationary solution $f(x,p)$ of Eq.\ (\ref{vlasov}) is the exponential
atmosphere:
\beq
f(x,p) = \exp \left[- \frac{\psi(x) - \psi(0)}{T_0} \right] f_0^{\rm M}(p)\, .
\label{ea}
\eeq
The system is isothermal, i.e., the temperature profile
\beq
T(x) = \frac{\frac{1}{2}\int_{-\infty}^\infty dp\, p^2 f(x,p)}{\int_{-\infty}^{\infty} dp\, f(x,p)} = \frac{1}{2 n(x)}\int_{-\infty}^\infty dp\, p^2 f(x,p)  
\eeq
is constant, $T(x) = T_0$. However, while velocity filtration occurs for
any boundary condition $f_0(p)$, it yields a constant temperature
profile only in the special case of the Maxwellian. This is due to the fact that $f(x,p)$ given by Eq.\ (\ref{ea}) is
separable in $x$ and $p$. In Ref.\ \cite{Scudder:apj1992a}, a graphical
argument was put forward as follows. If we plot $\ln f(x,p)$ as a
function of the kinetic energy $k = p^2/2$, we get a straight line. Now
$f(x,k)$ is obtained from $f(0,k)$ by removing the part of the
distribution corresponding to kinetic energies smaller than the
potential barrier $\Delta\psi(x)$, and then rigidly translating the remaining
part towards the origin. We thus get a straight line with the same
slope, and since in this case $T = - (d\ln f/dk)^{-1}$, the temperature
does not vary with $x$. 
However, the same graphical construction shows that if the tails of $f_0(p)$ are more populated than in a Maxwellian (i.e., $f_0(p)$ is suprathermal),
then velocity filtration yields a broader distribution at $x$, that is,
temperature increases as density decreases with increasing $x$. 
This qualitative argument is confirmed by direct calculations. To give an example, in Fig.\ \ref{fig_scudder_calculation} we plot the density and temperature profiles calculated solving Eq.\ (\ref{vlasov}) for a linear potential $\psi(x) = 25\,x$ and a power-law boundary condition
\beq
f_0(p) = \frac{\sqrt{2}}{\pi\left(1 + p^4 \right)}\,.
\label{f04}
\eeq

Why do we expect the same phenomenon to also occur in a
long-range-interacting system in an inhomogeneous QSS? Because for times
for which the system stays in a QSS, its one-particle distribution
function\footnote{Since we shall be mainly concerned with the HMF
model, we shall use $(\vartheta,p)$ as conjugated variables.} $f(\vartheta,p,t)$ still obeys a Vlasov equation,
\beq
\frac{\partial f}{\partial t} + p \frac{\partial f}{\partial \vartheta}
- \frac{\partial\left(\langle u \rangle + \psi\right)}{\partial \vartheta}\frac{\partial
f}{\partial p} = 0\, . 
\label{vlasov_lr}
\eeq 
The only difference with respect to Eq.\ (\ref{vlasov}) is that the
potential is now composed of two terms: the mean-field interaction,
\beq
\langle u \rangle (\vartheta,t) = \int d\vartheta' \int dp'\, u(\vartheta
- \vartheta') f(\vartheta',p',t), 
\label{u} 
\eeq
where $u$ is the two-body interaction, and (possibly) an external field $\psi(\vartheta)$, such that the total potential energy $V$ is 
\beq
V(\vartheta_1,\ldots,\vartheta_N) = \frac{1}{N}\sum_{i=1}^N \sum_{j < i}^N u(\vartheta_i - \vartheta_j)\, + \sum_{i=1}^N \psi(\vartheta)\,. 
\eeq
For the HMF model (\ref{hmf}), we have
\beq
u(\vartheta - \vartheta') = J \left[ 1 - \cos(\vartheta - \vartheta')\right]\, , 
\eeq
and
\beq
\psi(\vartheta) =  - h \cos \vartheta\,.
\eeq
The dynamics in a QSS is of Vlasov type because ``collisional'' effects
due to the finite number of particles become important only on a longer
time scale that diverges with $N$, and are those that drive the system
towards thermal equilibrium where $f(p)$ is Maxwellian
\cite{CampaEtAl:physrep}. Hence, despite the fact that each particle is
strongly interacting with all the other particles, the dynamics of our
system in a QSS is equivalent to that of noninteracting particles
immersed in a self-consistent field given by $\langle u \rangle$.
From the above discussion, it should be clear that the key ingredients
to obtain a temperature inversion due to velocity filtration in the
noninteracting particle case are $(i)$ the presence of an attractive
external field and $(ii)$ the fact that $f(p)$ is suprathermal. Hence,
we should expect the same behavior also in the long-range-interacting
case in a QSS, provided the resulting effective field $\langle u
\rangle$ is attractive---or, equivalently, the density profile is
clustered---and $f(p)$ is suprathermal. It is worth noting that our
situation is different from that of Ref.\ \cite{Scudder:apj1992a} not
only because the physical system under consideration is different, but
also that there the suprathermal momentum distribution is imposed as a
boundary condition, while in our case we shall only choose a
suprathermal initial condition for the dynamics. The initial momentum
distribution is expected to change under the dynamics, hence temperature
inversion may occur only if the suprathermal character of the initial
momentum distribution is not destroyed by the dynamical evolution. This
is to be checked a posteriori. 

\section{Simulations and results}
The MD simulations involved integrating the equations of motion
for the system (\ref{hmf}): 
\bea
\frac{d\vartheta_i}{dt} & = & p_i\, ,\\
\frac{dp_i}{dt} & = & J(m_y \cos
\vartheta_i-m_x\sin\vartheta_i)-h\sin
\theta_i\, ,
\label{hmf-eom}
\eea
where $i = 1,\ldots,N$ and
\bea
m_x & = &\frac{1}{N}\sum_{i=1}^N \cos \vartheta_i\, ,\\
m_y & =& \frac{1}{N}\sum_{i=1}^N \sin \vartheta_i\, .
\eea
We used a fourth-order symplectic integration algorithm
\cite{McLachlanAtela:nonlinearity1992}, with timestep $\Delta t =0.01$ that
ensured energy conservation up to $10^{-6}$. We simulated systems with $N = 5 \times 10^3$ particles and with $N = 10^5$ particles.
The initial state was prepared by either (i) putting all the particles at
$\vartheta=0$ (clustered initial state with $m=(m_x^2 + m_y^2)^{1/2} = 1$), or (ii) with
$\vartheta$'s uniformly distributed in $(0,2\pi]$ (homogeneous initial
state with $m=0$). The initial momenta
were sampled according to the ``kappa'' distribution \cite{Scudder:apj1992a}
\begin{equation}
P_\kappa(p)=\frac{\sqrt{\kappa}\Gamma(\kappa)}{w \sqrt{\pi}\Gamma(\kappa+1/2)}\left[1+\frac{p^2}{\kappa w^2}\right]^{-(\kappa+1)}\,
\label{kappa}
\end{equation}
that has power-law tails $\left|p\right|^{-2(\kappa+1)}$ and where $w^2$ is an effective temperature, becoming the usual one when $\kappa \to\infty$ so that $P_\kappa(p)$ converges to a Maxwellian. 
We let the system evolve for a sufficiently long time that it reaches a
QSS, and then collected data for the coordinates and the
momenta of the particles at different times separated by a decorrelation
time window. 

\begin{figure}
\centerline{\includegraphics[width=85mm]{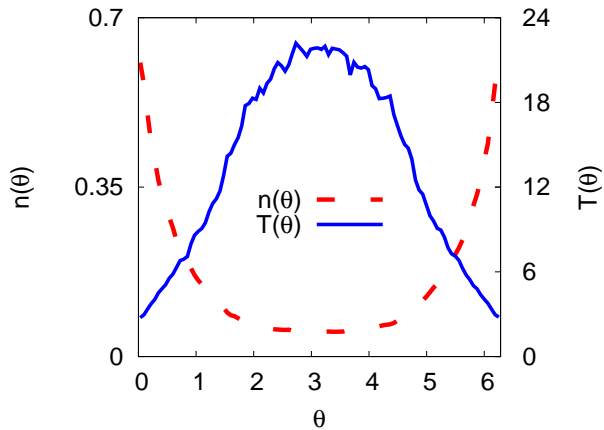}}
\caption{(Color online) AF case with $h=5$, $J=-1$ and $N = 5 \times 10^3$. Temperature (blue solid line) and density (red dashed line) profiles after the system has reached a QSS while starting from an initial state with $m=1$, $\kappa=1$ and $w=5$.}
\label{fig_AF_n_T}
\end{figure}
\begin{figure}
\centerline{\includegraphics[width=85mm]{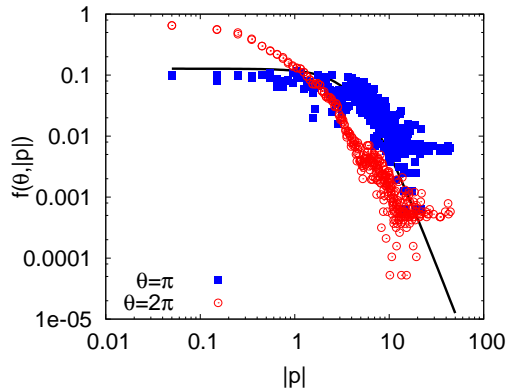}}
\caption{(Color online) AF case. Momentum distributions $f(\vartheta,|p|)$ in the same
conditions as in Fig.\ \protect\ref{fig_AF_n_T}, at $\vartheta=\pi$ (blue filled squares)
and at $\vartheta-2\pi$ (red open circles).
The black line is the initial momentum distribution (\ref{kappa}).}
\label{fig_AF_momenta}
\end{figure}
Let us start by presenting the results for the AF case, i.e., $J = -1$
and $h >0$ in Eq.\ (\ref{hmf}). Figure \ref{fig_AF_n_T} shows a typical
outcome for the density and temperature profiles when a system with $N = 5 \times 10^3$ particles has
reached a QSS after starting from the $m=1$ initial condition with $\kappa = 1$ and $w =
5$. The external field strength is $h=5$. Temperature
inversion is well evident, since temperature and density profiles are
anticorrelated. Figure \ref{fig_AF_momenta} shows the momentum
distribution $f(\vartheta,|p|)$ measured in the QSS at two different
positions, $\vartheta = 2\pi$ (i.e., the
highest-density point) and $\vartheta = \pi$ (i.e., the lowest-density
point). It is clear that there is a velocity filtration effect, since
$f(\pi,|p|)$ is much broader than $f(0,|p|)$. Moreover, although the
momentum distribution changes with respect to the initial one, the
suprathermal tails survive in the QSS. 

\begin{figure}
\centerline{\includegraphics[width=85mm]{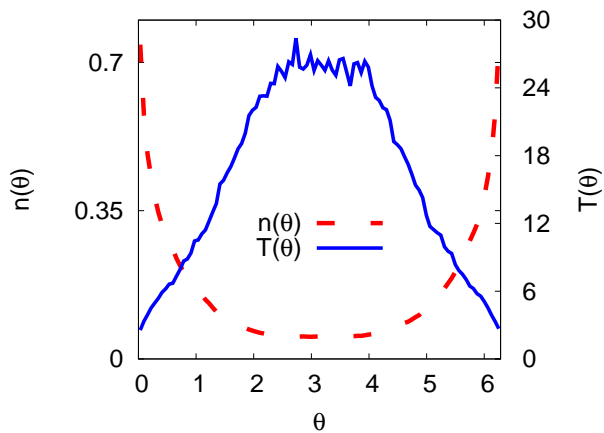}}
\caption{(Color online) As in Fig.\ \protect\ref{fig_AF_n_T}, with $N = 10^5$.}
\label{fig_AF_n_T_large}
\end{figure}
\begin{figure}
\centerline{\includegraphics[width=85mm]{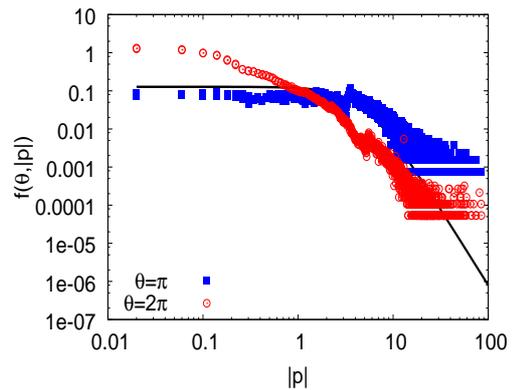}}
\caption{(Color online) AF case. Momentum distributions $f(\vartheta,|p|)$ in the same
conditions as in Fig.\ \protect\ref{fig_AF_n_T_large}, that is, with $N = 10^5$, at $\vartheta=\pi$ (blue filled squares)
and at $\vartheta-2\pi$ (red open circles).
The black line is the initial momentum distribution (\ref{kappa}).}
\label{fig_AF_momenta_large}
\end{figure}
Figures \ref{fig_AF_n_T_large} and \ref{fig_AF_momenta_large} show the density and temperature profiles and the momentum distributions, respectively, for a system with $N = 10^5$ particles, in the same conditions as in Figs.\ \ref{fig_AF_n_T} and \ref{fig_AF_momenta}. It is apparent that increasing $N$ does not change the phenomenology. 
  
We always observed temperature
inversion in the AF case, even starting from the $m=0$ initial
condition and varying $\kappa$ and $w$ in the initial momentum
distribution or varying the external field $h$ (data not shown). The
only changes observed were quantitative, i.e., in the amount of spatial
clustering in the QSS and in the strength of the temperature inversion.
Conversely, and as expected, we did not observe temperature inversion
when starting with a Maxwellian momentum distribution (data not shown). 

\begin{figure}
\centerline{\includegraphics[width=85mm]{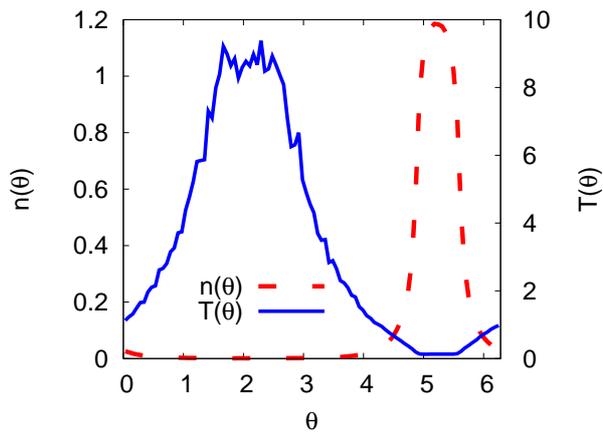}}
\caption{(Color online) F case with $J = 5$ and $N = 5 \times 10^3$. Temperature (blue solid line) and density (red dashed line) profiles after the system has reached a QSS while starting from an initial state with $m=1$, $\kappa=1$ and $w=1$.}
\label{fig_F_n_T}
\end{figure}

\begin{figure}
\centerline{\includegraphics[width=85mm]{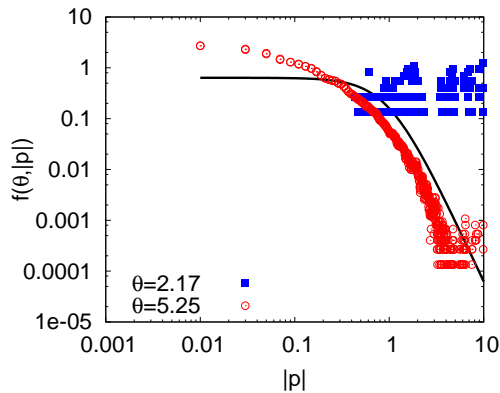}}
\caption{(Color online) F case. Momentum distributions $f(\vartheta,|p|)$ in the same
conditions as in Fig.\ \protect\ref{fig_F_n_T}, for 
$\vartheta= 2.17$ 
(blue filled squares) and  
$\vartheta = 5.25$ (red open circles). The
black line is the initial momentum distribution (\ref{kappa}).}
\label{fig_F_momenta}
\end{figure}

\begin{figure}
\centerline{\includegraphics[width=85mm]{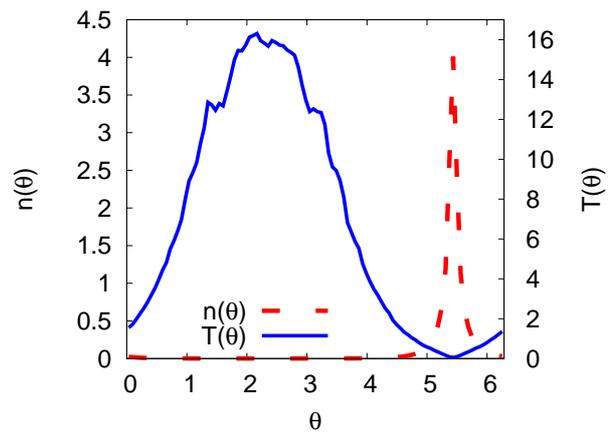}}
\caption{(Color online) As in Fig.\ \protect\ref{fig_F_n_T}, with $N =
10^5$, and $J=7$.}
\label{fig_F_n_T_large}
\end{figure}

\begin{figure}
\centerline{\includegraphics[width=85mm]{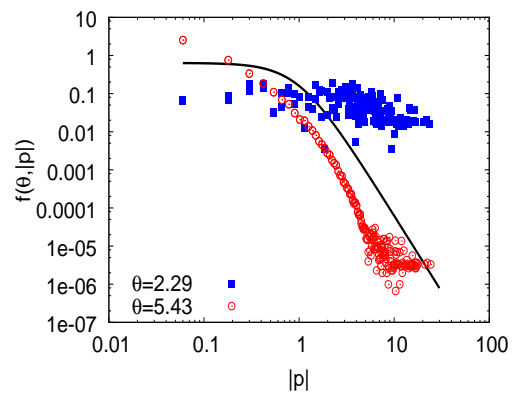}}
\caption{(Color online) F case. Momentum distributions $f(\vartheta,|p|)$ in the same
conditions as in Fig.\ \protect\ref{fig_F_n_T_large}, that is, with $N = 10^5$,  for 
$\vartheta= 2.29$ 
(blue filled squares) and  
$\vartheta = 5.43$ (red open circles). The
black line is the initial momentum distribution (\ref{kappa}).}
\label{fig_F_momenta_large}
\end{figure}

The phenomenology of the F case ($J >0$ and and $h = 0$) turns out to be
more complicated. The spatial structure of the QSS may vary a lot, from
a single cluster analogous to that always observed in the AF case to
multi-clustered structures with many superimposed density peaks to
almost uniform states, while starting even from very similar initial
conditions. This sensitive dependence on the choice of the initial
distribution is more pronounced for ``hotter'' initial data, i.e., with
$w \simeq 5$. For smaller values of $w$, one still observes either a
single cluster, or more than one cluster which tend to be close to
one another. The spatial structure of the QSS is thus strongly dependent
on the details of the initial conditions: this is a very interesting
phenomenon whose investigation is left for future work. This
notwithstanding, as regards temperature inversion, the situation is as
clear as in the AF case. Namely, whenever the QSS is appreciably clustered, we
do observe temperature inversion, even in multi-clustered states. An
example of density and temperature profiles for a QSS with a single
cluster, obtained setting $J=5$ and starting from the $m=1$ initial condition with
$\kappa = 1$ and $w = 1$ with $N = 5\times 10^3$ particles, is shown in Fig.\ \ref{fig_F_n_T}. In Fig.\
\ref{fig_F_momenta}, we show the momentum distribution at the highest
and lowest density points. Again, temperature inversion is well evident, and the velocity filtration effect is witnessed by the momentum distributions. Here the effect is particularly strong in that only the very fast particles survive in the low-density part of the system. 
As in the AF case, a similar situation can be observed also in a larger system. Density and temperature profiles for a QSS with a single
cluster, obtained setting $J=7$ and starting from the $m=1$ initial condition with
$\kappa = 1$ and $w = 1$ with $N = 10^5$ particles, is shown in Fig.\ \ref{fig_F_n_T_large}; the momentum distribution at the highest
and lowest density points is shown in Fig.\ \ref{fig_F_momenta_large}.

\section{Concluding remarks}
From our results, it is evident that nonthermal boundary conditions are
not necessary for temperature inversion due to velocity filtration. A
nonthermal initial condition, which could be modified by the violent
relaxation, may be sufficient, provided the initial suprathermal tails survive
and the stationary state reached by the system is sufficiently
clustered. The details of the spatial structure, however, are irrelevant
as to the occurrence of the temperature inversion effect. Temperature
inversion appears then as a very robust phenomenon, and it is reasonable
to expect that it may be present in any system with long-range interactions whenever the stationary state reached by the dynamical evolution is sufficiently clustered and has a suprathermal momentum distribution.

Clearly, we do not expect that every state resulting from violent relaxation exhibits a temperature inversion. For instance, in real self-gravitating systems like globular clusters the temperature profile is not flat, thus indicating that thermal equilibrium is not fully reached, but typically temperature decreases with the distance from the center \cite{BinneyTremaine:book,ZocchiBertinVarri:aa2012}. However, core-halo states \cite{LevinEtAlphysrep:arxiv2013} typically found in long-range-interacting systems support particle-wave interactions that may give high energy to halo particles and may thus support suprathermal distributions \cite{Teles:privcomm}; work is in progress along this direction. There are other astrophysical systems where anticorrelation between density and temperature profiles has been observed; an example is the hot gas in the so-called ``cooling-core'' galaxy clusters (see e.g.\ Refs.\ \cite{WiseMcNamaraMurray:apj2004,BaldiEtAl:apj2007}). Another example is provided by some dark molecular clouds \cite{PalmeirimEtAl:aa2013}.

It is interesting to note that a
non-uniform temperature profile was also found in a one-dimensional model of a driven
one-component plasma \cite{RizzatoPakterLevin:pre2009}, although in that case temperature and density profiles were directly correlated.

Since any distribution function that is a function of the one-particle
Hamiltonian is a  stationary solution  of the Vlasov equation
(\ref{vlasov_lr}), QSSs are naturally described as {\em stable}
stationary solutions. While many theoretical results are available for
the stability properties of spatially homogeneous solutions
\cite{CampaEtAl:physrep} in the HMF case, less is known in general on inhomogeneous solutions, despite some recent progress
\cite{CampaChavanis:jstat2010,BarreYamaguchi:arxiv2013}. Power-law
momentum distributions are stable for sufficiently high
energies in the homogeneous case \cite{CampaEtAl:physrep}. Our results
indicate that suprathermal tails of the momentum distribution may be generically stable features of the Vlasov dynamics also in the inhomogeneous case.

As previously stated, the aim of the present work was not to discuss
temperature inversion in its original astrophysics context as introduced
by Scudder \cite{Scudder:apj1992a,Scudder:apj1992b,Scudder:apj1994},
hence our results are not expected to be directly relevant to the coronal heating
problem. This notwithstanding, our results may suggest that postulating
a mechanism able to maintain a suprathermal distribution of the
velocities in the transition region below the solar corona may not be a
necessary condition to have velocity filtration and, as a consequence,
temperature inversion in the corona. A suprathermal velocity
distribution might simply be the effect of initial conditions that are
then dynamically self-sustained, provided the coronal plasma is in a QSS
which is natural to expect since the most
important interactions are the long-range ones. Let us emphasize that
temperature inversion cannot occur in systems with only short-range
interactions, since they do not have a Vlasov-like dynamics and
therefore do not generically sustain a QSS in which velocity distributions have suprathermal tails, a feature we pointed out to be crucial to obtaining temperature
inversion. 

\begin{acknowledgement}
We address a warm acknowledgment to Cesare Nardini for many fruitful discussions in the early stages of this work and for his interest and support.
SG acknowledges the support of the Indo-French Centre for the Promotion of Advanced
Research under Project 4604-3 and the hospitality of the Universit\`{a}
di Firenze.
\end{acknowledgement}

\bibliography{temp_inv}

\begin{thebibliography}{27}

\bibitem{Scudder:apj1992a}
J.D. Scudder, The Astrophysical Journal \textbf{398}, 299 (1992)

\bibitem{Scudder:apj1992b}
J.D. Scudder, The Astrophysical Journal \textbf{398}, 319 (1992)

\bibitem{Scudder:apj1994}
J.D. Scudder, The Astrophysical Journal \textbf{427}, 446 (1994)

\bibitem{MeyerVernetMoncuquetHoang:icarus1995}
N.~Meyer-Vernet, M.~Moncuquet, S.~Hoang, Icarus \textbf{116}, 202 (1995)

\bibitem{GolubPasachoff:book}
L.~Golub, J.M. Pasachoff, \emph{The Solar Corona}, 2nd~edn. (Cambridge
  University Press, Cambridge, 2009)

\bibitem{Aschwanden:book}
M.J. Aschwanden, \emph{Physics of the Solar Corona. An Introduction with
  Problems and Solutions}, 2nd~edn. (Springer, New York, 2005)

\bibitem{Klimchuk:sp2006}
J.A. Klimchuk, Solar Physics \textbf{234}, 41 (2006)

\bibitem{Anderson:apj1994}
S.W. Anderson, The Astrophysical Journal \textbf{437}, 860 (1994)

\bibitem{LandiPantellini:aa2001}
S.~Landi, F.G.E. Pantellini, Astronomy \& Astrophysics \textbf{372}, 686 (2001)

\bibitem{CampaEtAl:physrep}
A.~Campa, T.~Dauxois, S.~Ruffo, Physics Reports \textbf{480}(3-6), 57 (2009)

\bibitem{BouchetGuptaMukamel:physicaA2010}
F.~Bouchet, S.~Gupta, D.~Mukamel, Physica A: Statistical Mechanics and its
  Applications \textbf{389}(20), 4389  (2010), proceedings of the 12th
  International Summer School on Fundamental Problems in Statistical Physics

\bibitem{AntoniRuffo:pre1995}
M.~Antoni, S.~Ruffo, Phys. Rev. E \textbf{52}(3), 2361 (1995)

\bibitem{ChavanisVattevilleBouchet:epjb2005}
P.H. Chavanis, J.~Vatteville, F.~Bouchet, The European Physical Journal B -
  Condensed Matter and Complex Systems \textbf{46}, 61 (2005)

\bibitem{SotaEtAl:pre2001}
Y.~Sota, O.~Iguchi, M.~Morikawa, T.~Tatekawa, K.i. Maeda, Phys. Rev. E
  \textbf{64}(5), 056133 (2001)

\bibitem{TatekawaEtAl:pre2005}
T.~Tatekawa, F.~Bouchet, T.~Dauxois, S.~Ruffo, Phys. Rev. E \textbf{71}(5),
  056111 (2005)

\bibitem{jstat2010}
L.~Casetti, C.~Nardini, Journal of Statistical Mechanics: Theory and Experiment
  \textbf{2010}(05), P05006 (2010)

\bibitem{McLachlanAtela:nonlinearity1992}
R.I. McLachlan, P.~Atela, Nonlinearity \textbf{5}(2), 541 (1992)

\bibitem{BinneyTremaine:book}
J.~Binney, S.~Tremaine, \emph{Galactic Dynamics}, 2nd~edn. (Princeton
  University Press, Princeton, 2008)

\bibitem{ZocchiBertinVarri:aa2012}
A.~Zocchi, G.~Bertin, A.L. Varri, Astronomy \& Astrophysics \textbf{539}, A65
  (2012)

\bibitem{LevinEtAlphysrep:arxiv2013}
Y.~Levin, R.~Pakter, F.B. Rizzato, T.N. Teles, F.P. da~C.~Benetti (2013),
  \texttt{arXiv:1310.1078}

\bibitem{Teles:privcomm}
T.N. Teles, private communication

\bibitem{WiseMcNamaraMurray:apj2004}
M.W. Wise, B.R. McNamara, S.S. Murray, The Astrophysical Journal
  \textbf{601}(1), 184 (2004)

\bibitem{BaldiEtAl:apj2007}
A.~Baldi, S.~Ettori, P.~Mazzotta, P.~Tozzi, S.~Borgani, The Astrophysical
  Journal \textbf{666}(2), 835 (2007)

\bibitem{PalmeirimEtAl:aa2013}
{Palmeirim, P.}, {Andr\'e, Ph.}, {Kirk, J.}, {Ward-Thompson, D.}, {Arzoumanian,
  D.}, {K\"onyves, V.}, {Didelon, P.}, {Schneider, N.}, {Benedettini, M.},
  {Bontemps, S.} et~al., Astronomy \& Astrophysics \textbf{550}, A38 (2013)

\bibitem{RizzatoPakterLevin:pre2009}
F.B. Rizzato, R.~Pakter, Y.~Levin, Phys. Rev. E \textbf{80}, 021109 (2009)

\bibitem{CampaChavanis:jstat2010}
A.~Campa, P.H. Chavanis, Journal of Statistical Mechanics: Theory and
  Experiment \textbf{2010}(06), P06001 (2010)

\bibitem{BarreYamaguchi:arxiv2013}
J.~Barr\'e, Y.Y. Yamaguchi (2013), \texttt{arXiv:1311.3182}

\end{thebibliography}


\end{document}